\documentclass[
	a4paper, 
	10pt, 
	unnumberedsections, 
	twoside, 
]{LTJournalArticle}

\usepackage{hyperref}
\usepackage{wrapfig}
\usepackage{caption}
\usepackage[table]{xcolor}
\usepackage{colortbl}

\runninghead{Predictability over Volatility} 

\footertext{Preprint} 

\definecolor{color1}{RGB}{47, 86, 233}     
\definecolor{color11}{RGB}{255, 0, 0}    
\definecolor{color2}{RGB}{51, 84, 228} 
\definecolor{color3}{RGB}{52, 84, 223} 
\definecolor{color4}{RGB}{57, 82, 223} 
\definecolor{color5}{RGB}{72, 82, 221} 
\definecolor{color6}{RGB}{73, 75, 204} 
\definecolor{color7}{RGB}{82, 71, 193} 
\definecolor{color8}{RGB}{122, 55, 149} 
\definecolor{color9}{RGB}{125, 54, 146} 
\definecolor{color10}{RGB}{220, 14, 40} 
\definecolor{color_missing1}{RGB}{103, 63, 170} 
\definecolor{lightblue}{rgb}{0.8, 0.88, 1.0}
\definecolor{darkblue}{rgb}{0, 0, 139}

\title{Predictability over Volatility: When Simple Strategies Help Households Save on Fuel} 

\author{%
	Shahrzad Jamshidi\textsuperscript{1}, \thanks{Corresponding author: \href{mailto:sjamshidi@lakeforest.edu}{sjamshidi@lakeforest.edu} This work was partially funded by the Richter Scholar Summer Research Program at Lake Forest College. The authors thank Nancy Tao, Dimitra Papadovasilaki, Tilahun Emiru and Linh Pham for valuable feedback on the project. We also thank Arthur Bousquet for assistance in generating Figure~\ref{fig:gas_model}. AI assistants were used in preparing this manuscript: Le Chat (Mistral AI), Claude (Anthropic), and Apple Intelligence (Apple) were used to help edit and refine drafts, and Adobe's generative AI tools were used in creating Figure~\ref{fig:gas_model}. The authors reviewed all AI-assisted output and take full responsibility for the content.}  \\
    Ako Ogihara\textsuperscript{2} \normalsize 
}

\date{\footnotesize\textsuperscript{\textbf{1}} Mathematics and Computer Science, Lake Forest College \\
\footnotesize\textsuperscript{\textbf{2}} Department of Economics, Business, and Finance, Lake Forest College}

\renewcommand{\maketitlehookd}{%
	\begin{abstract}
		\noindent What reasonable strategies can households use to save on gasoline or diesel purchases under different pricing functions? We assess basic strategies households might employ under a simplified model of a 10-gallon tank, fixed fuel consumption between refill opportunities, and price changes at each opportunity to refill. Purchases always fill the tank, and if the tank is empty, it must be refilled regardless of price. We explore three strategy types: price-independent (filling once the tank is below a specific level), price-dependent (filling when the price is below the last $n$ days or their average), and mixed (filling only below a set level, and then only when the price is below the last $n$ days or their average). We construct price functions from the nearest-in-time US weekly averages of regular gasoline and diesel, filling in higher-frequency values with a smoothing spline, linear interpolation, and random draws from normal and uniform distributions based on national averages. Additional simulations are built using Brownian motion. Together, these pricing models span a wide range of volatilities. We find that price-independent strategies are effective for the spline and linear functions, which have low volatility, while price-dependent strategies are effective for the normal and uniform distributions, which are highly volatile. Brownian motion, despite its mid-range volatility, admits no effective strategy. Our findings suggest that the predictability of price movements—whether prices trend or revert toward a stable level—matters more for household strategies than overall volatility: the random walk is volatile in the middle of our range yet offers households no usable signal. Policymakers can therefore assist households by addressing how prices move from day to day, not only the retail prices they face.
	\end{abstract}
}

\begin{document}

\maketitle 


\section{Introduction}

Gasoline is one of the largest recurring household expenditures: average annual household spending on gasoline was \$2{,}148 in 2021, surpassing combined spending on electricity, natural gas, and fuel oil according to \cite{BLS_CEX_2021}. Yet demand for fuel is relatively price-inelastic; recent estimates from \cite{Coglianese2017,LevinLewisWolak2017} place the short-run elasticity between roughly $-0.3$ and $-0.4$ once endogeneity and anticipatory purchasing are taken into account. Because quantity adjusts so little, a change in the price paid passes almost directly into the household budget: \cite{Gelman2023} find that the marginal propensity to consume out of gasoline-price windfalls is close to one, and \cite{HastingsShapiro2013} show that households do not even treat such windfalls as fully fungible, adjusting fuel-grade choice more than a standard income effect would predict. Small reductions in the average price a household pays are therefore economically meaningful. But precisely because households cannot easily consume less fuel, the lever available to them is not how much they buy but \emph{when} and \emph{how} they buy it. This is the question this paper explores.

What a purchase-timing decision can accomplish depends on how retail prices actually move. Retail gasoline prices are driven primarily by crude oil costs, but the pass-through is neither immediate nor symmetric: prices rise faster than they fall. This is sometimes referred to as the ``rockets and feathers'' asymmetry, first named and quantified by \cite{Bacon1991} and later documented for US retail markets by \cite{Borenstein1997}. In many markets, they trace pronounced Edgeworth cycles of sharp increases followed by slow undercutting \cite{Noel2007}. Surveying this literature, \cite{Eckert2013} emphasizes that local market structure and retailer behavior, not crude costs alone, govern the timing and size of retail price adjustments. The upshot is that retail prices are persistent, serially dependent, and often trending rather than pure random noise. Our synthetic price processes discussed in \nameref{sec:methods} span a range of noisiness for this reason.

How might a household exploit such structure? A long line of work including  \cite{HendelNevo2006,HendelNevo2013} shows that consumers shift purchases of storable goods toward temporarily low prices. According to \cite{KalyanaramWiner1995,Mazumdar2005}, they judge prices not in absolute terms but against an internal reference formed largely from recently observed prices. Households are thus likely to rely on simple, history-based rules, such as the lowest price in X days, rather than explicit forecasts. This aligns well with research on bounded rationality since fast, low-information heuristics can perform well relative to fully optimizing rules in uncertain environments \cite{Simon1955,GigerenzerGoldstein1996}. The strategies we study, discussed in \nameref{sec:strategies}, are heuristics of exactly this kind, requiring little information and no forecasting.

Whether any such rule can help, however, depends entirely on how predictable prices are and predictability must be distinguished from volatility. If prices fully reflect available information, price changes are essentially unforecastable according to \cite{Samuelson1965}. This is the random-walk benchmark discussed by \cite{Fama1970}, under which past prices carry no exploitable information about future ones. Real series depart from that benchmark to varying, measurable degrees (see \cite{LoMacKinlay1988}), yet for energy prices the no-change (random-walk) forecast is notoriously hard to beat, as discussed by \cite{AlquistKilian2010}. Volatility is a separate matter: the ARCH and GARCH models of \cite{Engle1982} and \cite{Bollerslev1986} describe how the \emph{size} of price movements clusters and persists, but say nothing about their \emph{direction}. A price process can therefore be highly volatile yet directionally unpredictable. This distinction between predictable magnitude and predictable direction is important to this study.

\begin{figure} 
    \begin{center}
	\includegraphics[width=0.6\linewidth]{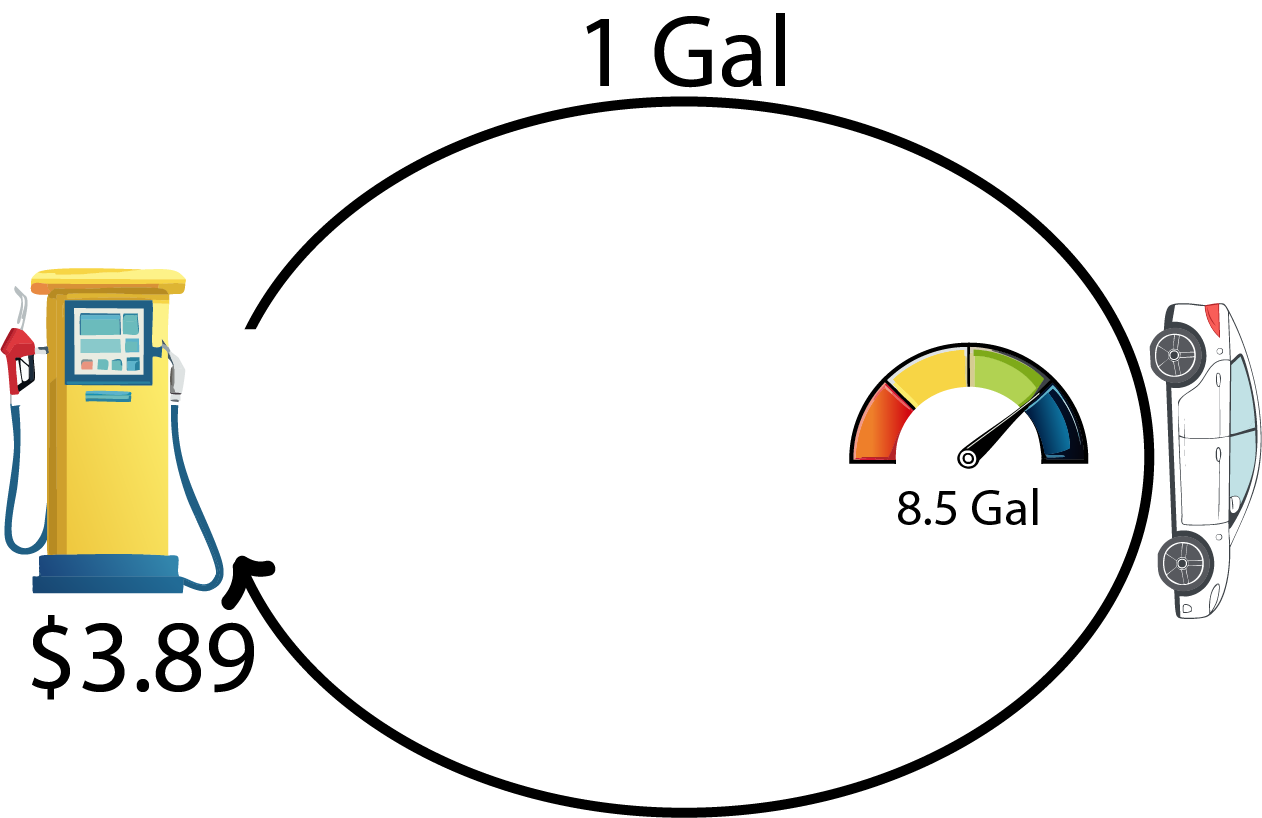}
    \end{center}
	\caption{The model simplifies fuel purchasing by assuming a 10 gallon fuel tank and a consumption rate of 1 gallon of fuel between refueling opportunities. If a consumer elects to refuel, the tank is filled to capacity. If the tank is empty, it must be refilled regardless of cost. }
	\label{fig:gas_model}
\end{figure}

This paper asks which simple household purchasing rule, if any, pays off under which kind of price process. We model fuel purchasing deliberately simply (Figure~\ref{fig:gas_model}): a household with a ten-gallon tank consumes one gallon between refueling opportunities and, whenever it chooses to refuel, fills to capacity. If the tank runs empty, it must be refilled regardless of price. The figures of ten and 1 were taken based on order of magnitude estimates of gas-tank sizes, which vary widely, and on estimates of gas station distribution. 

Holding this model and the space of purchasing rules fixed, we vary the price process across a wide range of volatilities, from smooth trends to pure random walks, and evaluate each rule against a random-purchasing benchmark. Our central finding is that it is the predictability of price movements, not their volatility, that determines whether a rule beats random buying: the random walk is volatile and so offers no usable signal and so no strategy is successful. Yet other price processes have usable strategies that either respond to price or are mixtures of price responses and regular purchasing. Because real gasoline and diesel prices are themselves trending and autocorrelated, that simple rule is enough: households need not track prices long-term to capture the available savings. 

The remainder of the paper is organized as follows: \nameref{sec:methods} describes the model, price processes, and strategy space; \nameref{sec:analysis} reports results by price regime; and \nameref{sec:discussion} interprets them and draws out the policy implication.

\section{Methods}\label{sec:methods}

To compare refueling strategies, this study simplifies the real world to a model in Figure \ref{fig:gas_model}. We assume a single fuel supplier, a 1-gallon consumption rate between refueling opportunities, and a 10-gallon tank capacity. Prices adjust between refueling opportunities, and consumers must refill the tank when empty but can top off sooner. Put abstractly, consumers must maintain a target commodity stock (with a maximum capacity), have regular purchasing and consumption, and must buy when depleted. While we discuss gasoline and diesel purchases, this approach could apply to other inelastic necessities.

The simplified model is based on order-of-magnitude calculations, not actual fuel tank capacities or station distributions. Top-selling gas-powered cars of 2024, as listed in \cite{Ireson2025}, have models like Ford F-150s with 36-gallon tanks (\cite{Ford2024}) and Hondas with 12.4-gallon tanks (\cite{Honda2024}). Their orders of magnitude are the same (10), so we construct the model from Fermi estimates. The thinking is similar with respect to fuel consumption between refueling opportunities. 

\subsection{Pricing Functions}\label{sec:datasets}

Gasoline and diesel prices are influenced by factors like crude oil, refining, transportation, marketing, and more. We construct price models based on weekly average U.S. gasoline and diesel prices between April 2022 and February 2025 calculated by the American Automobile Association and sourced from Statista (\cite{AAA2025}) to account for some of these factors.  
The price models used in this study are:

\begin{figure} 
    \centering
	\includegraphics[width=0.8\linewidth]{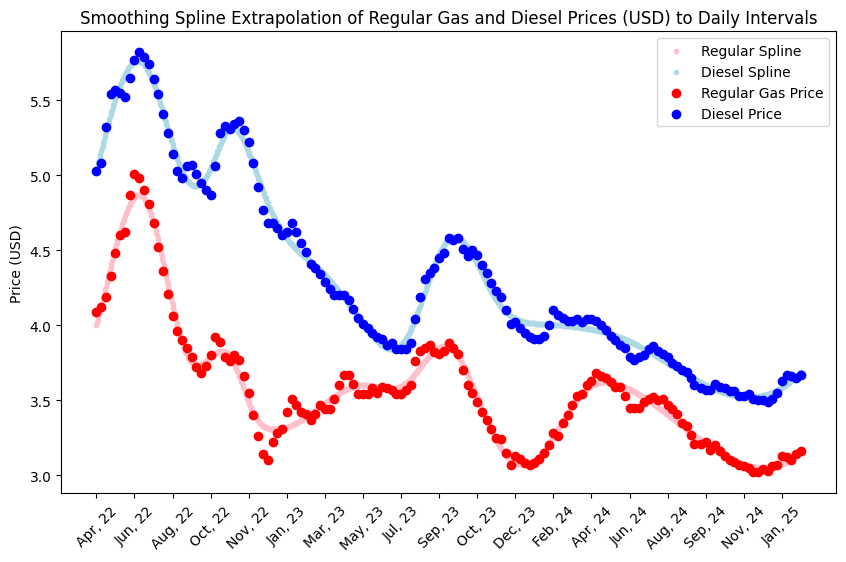}
	\caption{The data depicted is regular gas and diesel from \cite{AAA2025}. The smoothing spline extrapolation is used to simulate hypothetical gas price data with extremely low volatility.}
	\label{fig:smoothing_spline}
\end{figure}

\begin{figure*} 
\centering
\includegraphics[width=\linewidth]{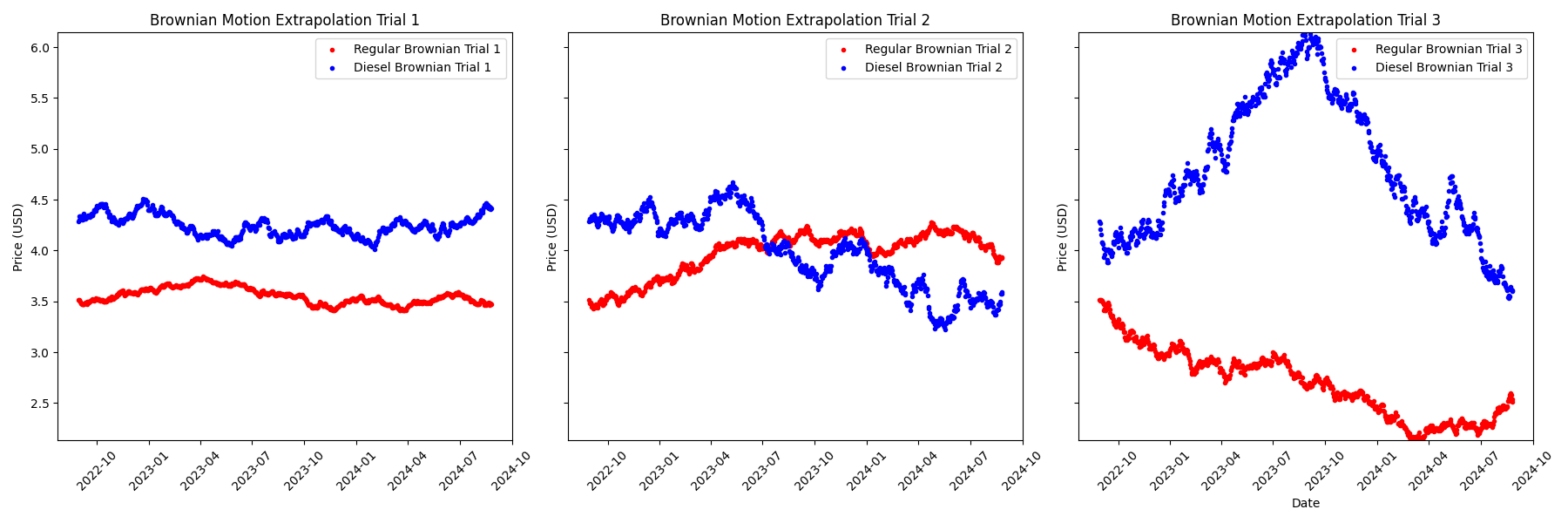}
	\caption{Three Brownian trials are introduced with varying volatilities (listed in order from left to right). These trials exhibit similar statistical behavior from \cite{AAA2025} but follow very different trends.}
	\label{fig:brownian}
\end{figure*}

\begin{itemize}
    \item \textbf{Smoothing Spline:} Smoothing splines are flexible curves fitted to data points that balance fidelity to the data with the smoothness of the curve, minimizing a combination of the residual sum of squares and the integrated square of the second derivative. This price model reflects fuel prices that are very stable but trending, as depicted in Figure \ref{fig:smoothing_spline}. Short-term fluctuations and noise are absent from this model, allowing us to assess how consumers can respond to trend alone. 
    \item \textbf{Linear Interpolation:} This model essentially "connects the dots" in our dataset. It contains gradual changes with trend, similar to the smoothing spline model, but adjusts sharply on a weekly basis. 
    \item \textbf{Brownian Motion (Random Walk):} This model reflects a market of high volatility and uncertainty, absent of historical influence. It is useful for simulating scenarios where prices fluctuate randomly, providing insights into how consumers might respond to unpredictable price changes. We generated four Brownian datasets based on the means and variances of REG and DSL in \cite{AAA2025}, each with increasing variances. Figure \ref{fig:brownian} illustrates these three trials. 
    \item \textbf{Random Draw from a Normal Distribution:} This model simulates price changes that cluster around a mean value constructed from the closest points in time from \cite{AAA2025}, with the range defined by the standard deviation of the same points. Here, the noise is present but somewhat controlled and the overall trend is still present. The normal distribution model accounts for noise and random fluctuations that could be indicative of many gas station prices within a region, providing perhaps the most realistic representation of price changes in the US market faced by consumers.
    \item \textbf{Random Draw from a Uniform Distribution:} This model ensures that all price points within a specified range defined by the closest points in time from \cite{AAA2025} have an equal probability of occurring. Here, the volatility is quite high but we still have an overall trend and historical price influence.  
\end{itemize} 
These 16 constructed price models (6 are random walks, and 4 are uniform) have varying volatilities, listed in Table \ref{tab:spectrum}. Each has 729 points, arising from the 105 samples present in \cite{AAA2025}.

\begin{table}
    \centering
    \resizebox{0.45\textwidth}{!}{
        \begin{tabular}{|c|c|l|}
            \hline
            \cellcolor{black}\textcolor{white}{\bf REG Vol.} & \cellcolor{black}\textcolor{white}{\bf DSL Vol.} & \cellcolor{black}\textcolor{white}{\bf Data Generation Method} \\ \hline
            \cellcolor{color1} \textcolor{white}{0.0017} & \cellcolor{color1} \textcolor{white}{0.0014} & Smoothing Spline \\ \hline 
            \cellcolor{color1} \textcolor{white}{0.0026} & \cellcolor{color1} \textcolor{white}{0.0020} & Linear \\ \hline 
            \cellcolor{color2} \textcolor{white}{0.0080} & \cellcolor{color3} \textcolor{white}{0.0092} &  Brownian T1 \\ \hline 
            \cellcolor{color4} \textcolor{white}{0.0137} & \cellcolor{color5} \textcolor{white}{0.0127} & Brownian T2 \\ \hline \hline 
            \cellcolor{color7} \textcolor{white}{0.0210} & \cellcolor{color6} \textcolor{white}{0.0154} & Brownian T3 \\ \hline 
            \cellcolor{color7} \textcolor{white}{0.0213} & \cellcolor{color6} \textcolor{white}{0.0177} & Normal Sampling (3 wks) \\ \hline 
            \cellcolor{color8} \textcolor{white}{0.0302} & \cellcolor{color9} \textcolor{white}{0.0251} & Uniform Sampling (3 wks) \\ \hline 
            \cellcolor{color11} \textcolor{white}{0.0494} & \cellcolor{color10} \textcolor{white}{0.0377} & Uniform Sampling (5 wks) \\ \hline 
        \end{tabular}
    }
    \caption{Generated datasets ranked by the statistical volatility, which we calculate to be the (non-annualized) standard deviation of the percent change, using a daily timescale. Cell coloring is a visual indicator of scale. Weekly Statistical volatility observed from \cite{AAA2025} are 0.0172 (REG) and 0.0135 (DSL). The table split indicates their  location. }\label{tab:spectrum}
\end{table}

\begin{table*}[h!]
\centering
\begin{tabular}{lccccccc}
\toprule
 & \multicolumn{7}{c}{Lag Coefficients} \\
\cmidrule(l){2-8}
Fuel Type & Lag 1 & Lag 2 & Lag 3 & Lag 4 & Lag 5 & Lag 6 & Lag 7 \\
\midrule
Regular & 0.235 & 1.542 & -0.659 & 0.131 & -0.083 & -- & -- \\
Diesel & 0.096 & 1.731 & -0.952 & 0.332 & -0.156 & 0.094 & -0.164 \\
\bottomrule
\end{tabular}
\caption{Autoregressive (AR) coefficients for weekly prices of Regular (REG) and Diesel (DSL) with 5 and 7 lags respectively.}
\label{tab:coefficients}
\end{table*}

The mean absolute weekly changes in prices within the original data are observed to be \$0.049 (1.3850\%) and \$0.046 (1.0504\%) for Regular (REG) and Diesel (DSL), respectively. The non-annualized statistical volatilities, measured as the standard deviation of the weekly percent change, are 0.0172 for REG and 0.0135 for DSL. Both REG and DSL exhibit significant heteroskedasticity and autocorrelation. For REG, the Breusch-Pagan test indicated non-constant variance in residuals, with a \( p \)-value of \( 1.695 \times 10^{-5} \). Additionally, the Durbin-Watson statistic of 0.0249 and the Ljung-Box test (\( p \)-value: \( 2.345 \times 10^{-176} \)) confirmed strong positive autocorrelation. The ACF plot suggests significant lag coefficients up to 5 weeks. Similarly, DSL showed significant heteroskedasticity, as evidenced by the Breusch-Pagan test (\( p \)-value: \( 4.007 \times 10^{-6} \)). The Durbin-Watson test statistic of 0.0110 also indicated strong positive autocorrelation, further supported by the Ljung-Box test (\( p \)-value: \( 8.441 \times 10^{-237} \)). The ACF plot for DSL suggests an autocorrelation extending up to 7 weeks. Table~\ref{tab:coefficients} displays the autoregressive (AR) coefficients for time series data for both REG and DSL. 

\subsubsection{Comparisons.}  For a comparative analysis using weekly data over the same time period, three key benchmarks were considered: Europe's Brent crude oil, the U.S. WTI crude oil, and OPEC's basket. According to \cite{Statista2025OPEC}, weekly prices from April 19, 2022, to February 12, 2025, exhibited non-annualized statistical volatilities of 0.0412, 0.0408, and 0.0466, respectively. These values indicate significantly greater price fluctuations compared to those reported in \cite{AAA2025}. Additionally, the mean absolute percent changes for these benchmarks were higher, with values of 3.3\%, 3.2\%, and 3.7\%, respectively, indicating larger week-to-week price fluctuations. In terms of autocorrelation, the patterns are relatively similar, with statistically significant correlations observed over the last 8 weeks.

We also performed comparisons with weekly average prices in the same date range for a set of countries, based on data availability.

Ghana exhibited higher statistical volatility (0.0766), with weekly regular gasoline prices from April 18, 2022, to February 17, 2025 taken from \cite{Statista2025Ghana}. The mean absolute percent change during this period was 2.0773\%, nearly double that of REG. This dataset exhibited autocorrelation of 5 weeks, slightly shorter, contributing to more instability.

Conversely, the UK experienced lower volatilities of 0.0091 (regular) and 0.0092 (diesel) over the same period taken from \cite{Statista2025UK}, suggesting more stability. The mean absolute percent changes were 0.6514\% and 0.7019\%, respectively. Both have longer autocorrelation lags of 10 and 11 weeks, respectively, further supporting higher stability in pricing.

C\^ote d'Ivoire (regular, \cite{Statista2025CoteDeIvoire}), France (diesel \& super 95, \cite{Statista2025France}), and Italy (super 95, \cite{Statista2025Italy}) experienced higher volatility comparable to the "Normal Sampling 3 wks" simulated dataset, with mean absolute percent changes ranging from 1.1\% to 1.3\%. They exhibited autocorrelation lags of 10 weeks, 5 \& 7 weeks, and 5 weeks, respectively. The longer autocorrelation lag in C\^ote d'Ivoire suggests a more extended influence of past prices, while the shorter lags in France and Italy indicate a more immediate impact.

None of the datasets exhibited stationarity and all showed signs of heteroskedasticity, with seasonality potentially being a factor.

\subsection{Consumer Purchasing Strategies}\label{sec:strategies}

Household purchasing strategies vary. We opt for simple consumer strategies based on historical data tabulations. These strategies are
\begin{enumerate}
    \item {\bf Lower than the previous $n$ Days}: purchasing early when the price is the lowest in the last $n$ days, \label{lowest_n}
    \item {\bf Lower than the previous $n$-Day Average}: purchasing early when the price is lower than the average of the last $n$ days, \label{lowest_avg}
    \item {\bf Fill when Tank is Below $X\%$}: purchasing at a specified tank level, irrespective of price, and\label{levels}
    \item {\bf Strategic Purchasing when Tank is Below $X\%$}: mixture strategies that only consider purchasing below a specified tank level. \label{mixed}
\end{enumerate} We consider values of $n$ ranging from 1 to 30, approximately one month. The tank levels considered are zero, one-tenth, one-quarter, one-third, and one-half. Notice that these regular refuelings are equivalent to refueling every 10, 9, 8, 7, and 5 days, respectively. For this reason, we also looked at lags up to 6 days to account for the weekly nature of the original data. As a result, we explore 1,775 strategies total.

These strategies were selected since they were simple and fall nicely into price-based (\ref{lowest_n}, \ref{lowest_avg}), need-based (\ref{levels}) and mixed strategies (\ref{mixed}). The use of simpler mathematical functions based on the recent price history can be justified here by the concepts of bounded rationality and limited cognitive energy. Instead of complex optimization models that assume perfect information and unlimited computational capacity, consumers often rely on heuristics and rules of thumb to simplify decision-making processes, primarily because they face numerous decisions daily. Simpler strategies may be more practical, as the potential annual savings from strategic gas purchasing may not justify constant tracking and modeling. 

\subsection{Random Purchasing Distributions} 

For each of the 16 datasets presented in Table \ref{tab:spectrum}, we generated a reference distribution comprising 1000 average prices from randomized purchases. A random purchasing strategy uniformly selects a day at random from the subsequent 10-day period after the last purchase; the tank is then topped off on that chosen day. Notice that the random purchase strategy conforms to the model parameters (see Figure \ref{fig:gas_model}). 

Given that these random distributions are based on the average prices per gallon across 1000 distinct trials, the Central Limit Theorem is applicable. The resulting distributions are approximately normal, something confirmed by the Shapiro-Wilk test and the Kolmogorov-Smirnov test. Distribution data is available in Table \ref{tab:stats}.

For each strategy, we compare the average price per gallon to the reference distribution to compute the probability of obtaining a similar average price at random. Within each simulation, we identify strategies that beat 95\% or more of the random purchasing out of the 1,775 strategies. These strategies consider three tank levels (0\%, 10\%, and 25\%), 30 potential values of $n$, and 7 possible lags (0 through 6).

We naturally group these individual strategies and datasets. We focus on the overall performance of strategy categories within each simulation type to say, in general, this type of strategy appears to translate into savings compared to random purchasing. 

\begin{table*}
\centering
\begin{tabular}{l|ccccc}
\toprule
\rowcolor{darkblue} 
\textcolor{white}{Simulation Name} & \textcolor{white}{Mean} & \textcolor{white}{Std. Dev.} & \textcolor{white}{Range Interval} & \textcolor{white}{Skewness} & \textcolor{white}{Kurtosis} \\
REG Spline & 3.568195 & 0.001145 & (3.571531, 3.564463) & -0.271530 & -0.375846 \\
\rowcolor{lightblue}
REG Linear & 3.567642 & 0.001343 & (3.571249, 3.563194) & -0.113646 & -0.183898 \\
REG Brownian T1 & 3.325045 & 0.002250 & (3.331961, 3.318568) & 0.225481 & 0.046182 \\
\rowcolor{lightblue}
REG Brownian T2 & 3.362894 & 0.003422 & (3.373948, 3.350751) & 0.068575 & -0.111659 \\
REG Brownian T3 & 3.632557 & 0.006167 & (3.649018, 3.612742) & -0.170661 & -0.163563 \\
\rowcolor{lightblue}
REG Normal 3wk & 3.531172 & 0.004529 & (3.548024, 3.517642) & 0.170056 & 0.037872 \\
REG Uniform 3 wk & 3.565343 & 0.006632 & (3.589357, 3.547949) & 0.177893 & -0.095180 \\
\rowcolor{lightblue}
REG Uniform 5 wk & 3.562967 & 0.010589 & (3.597776, 3.525721) & -0.075205 & -0.063366 \\
DSL Spline & 4.316215 & 0.001606 & (4.319709, 4.311843) & -0.395817 & -0.581205 \\
\rowcolor{lightblue}
DSL Linear & 4.316222 & 0.001788 & (4.320526, 4.311326) & -0.270612 & -0.609729 \\
DSL Brownian T1 & 4.259859 & 0.003181 & (4.269291, 4.250644) & -0.001547 & -0.195492 \\
\rowcolor{lightblue}
DSL Brownian T2 & 5.294450 & 0.004918 & (5.310694, 5.280542) & 0.064332 & -0.003215 \\
DSL Brownian T3 & 6.670360 & 0.007575 & (6.697757, 6.642411) & 0.003721 & 0.113541 \\
\rowcolor{lightblue}
DSL Uniform 3 wk & 4.315124 & 0.006650 & (4.333889, 4.296002) & -0.112448 & -0.248689 \\
DSL Normal 3 wk & 4.284526 & 0.004440 & (4.298034, 4.269051) & -0.106373 & 0.083358 \\
\rowcolor{lightblue}
DSL Uniform 5 wk & 4.315570 & 0.009362 & (4.347989, 4.281421) & 0.012853 & 0.052776 \\
\bottomrule
\end{tabular}
\caption{Descriptive statistics of average price per gallon from 1000 trials of random purchases for each simulated dataset. These distributions are approximately normal.}
\label{tab:stats}
\end{table*}

\section{Simulations and Analysis}\label{sec:analysis}

\begin{table}[h!]
\centering\scriptsize
\begin{tabular}{@{\,}l@{\,}l@{\,}l@{\,}l@{\,}}
\toprule
\textbf{Dataset} & \textbf{Strategy} & \textit{Avg?} & \textit{Cases} \\ \midrule \hline
Spline   & Fill at & N/A & 0\%, 10\%, 33\%, 50\% \\ \cline{2-4}
\& Linear          & Price Dependent & Y & \cellcolor{gray!20}None\\ \cline{3-4}
                & & N & \cellcolor{gray!20}None \\ \cline{2-4}
                & Mixed & Y & 1-4, 7, 20-22, 25 days at 10\%\\ \cline{3-4}
                & & N & 1-3, 12-22 days at 10\%\\ \hline \midrule \hline
Brownian & Fill at & N/A & \cellcolor{gray!20}None \\ \cline{2-4}
                & Price Dependent & Y & \cellcolor{gray!20}None\\ \cline{3-4}
                & & N & \cellcolor{gray!20}None \\ \cline{2-4} 
                & Mixed & Y & \cellcolor{gray!20}None \\ \cline{3-4}
                & & N & \cellcolor{gray!20}None\\ \midrule \hline
Normal   & Fill at & N/A & 0\%\\ \cline{2-4}
\& Uniform         & Price Dependent & Y & 1-13 days\\ \cline{3-4}
                & & N & \cellcolor{yellow}All \\ \cline{2-4}
                & Mixed & Y & \cellcolor{yellow}All \\ \cline{3-4}
                & & N & \cellcolor{yellow}All \\ \hline \bottomrule
\end{tabular}
\caption{Strategies that performed better than 95\% of random purchases for all simulations of the specified type.}
\label{tab:performance}
\end{table}\normalsize


\begin{table*}[h]
\centering
\setlength{\tabcolsep}{4pt} 
\small 
\begin{tabular}{l|cc|ccccc}
\toprule
\multicolumn{2}{c}{} & & \multicolumn{5}{c}{\textcolor{black}{Probabilities}} \\
\midrule
\textcolor{black}{Dataset} & \textcolor{black}{Mean} & \textcolor{black}{Std} & \textcolor{black}{Fill at $p$} & \textcolor{black}{$<n$-Day Avg} & \textcolor{black}{$<n$ Days} & \textcolor{black}{Mixed$<n$-Day Avg} & \textcolor{black}{Mixed < $n$ Days} \\
\midrule
{\it count} & - & - & {\it 35} & {\it 30} & {\it 30} & {\it 840} & {\it 840} \\ 
\midrule
REG Spline & 3.5682 & 0.0011 & \cellcolor{white}\textcolor{black}{0.0000} & \cellcolor{gray!50}\textcolor{black}{1.0000} & \cellcolor{gray!50}\textcolor{black}{1.0000} & \cellcolor{white}\textcolor{black}{0.0000} & \cellcolor{white}\textcolor{black}{0.0000} \\
REG Linear & 3.5676 & 0.0013 & \cellcolor{white}\textcolor{black}{0.0000} & \cellcolor{gray!50}\textcolor{black}{1.0000} & \cellcolor{gray!50}\textcolor{black}{1.0000} & \cellcolor{white}\textcolor{black}{0.0000} & \cellcolor{white}\textcolor{black}{0.0000} \\
REG Brownian T1 & 3.3250 & 0.0023 & \cellcolor{white}\textcolor{black}{0.0000} & \cellcolor{white}\textcolor{black}{0.0099} & \cellcolor{gray!50}\textcolor{black}{0.3656} & \cellcolor{white}\textcolor{black}{0.0000} & \cellcolor{white}\textcolor{black}{0.0000} \\
REG Brownian T2 & 3.3629 & 0.0034 & \cellcolor{white}\textcolor{black}{0.0000} & \cellcolor{white}\textcolor{black}{0.0000} & \cellcolor{white}\textcolor{black}{0.0000} & \cellcolor{white}\textcolor{black}{0.0000} & \cellcolor{white}\textcolor{black}{0.0000} \\
REG Brownian T3 & 3.6326 & 0.0062 & \cellcolor{white}\textcolor{black}{0.0000} & \cellcolor{white}\textcolor{black}{0.0000} & \cellcolor{white}\textcolor{black}{0.0000} & \cellcolor{white}\textcolor{black}{0.0000} & \cellcolor{white}\textcolor{black}{0.0000} \\
REG Normal & 3.5312 & 0.0045 & \cellcolor{white}\textcolor{black}{0.0000} & \cellcolor{white}\textcolor{black}{0.0000} & \cellcolor{white}\textcolor{black}{0.0000} & \cellcolor{white}\textcolor{black}{0.0000} & \cellcolor{white}\textcolor{black}{0.0000} \\
REG Uniform 3 & 3.5653 & 0.0066 & \cellcolor{white}\textcolor{black}{0.0000} & \cellcolor{white}\textcolor{black}{0.0000} & \cellcolor{white}\textcolor{black}{0.0000} & \cellcolor{white}\textcolor{black}{0.0000} & \cellcolor{white}\textcolor{black}{0.0000} \\
REG Uniform 5 & 3.5630 & 0.0106 & \cellcolor{white}\textcolor{black}{0.0000} & \cellcolor{white}\textcolor{black}{0.0000} & \cellcolor{white}\textcolor{black}{0.0000} & \cellcolor{white}\textcolor{black}{0.0000} & \cellcolor{white}\textcolor{black}{0.0000} \\
 \hline \hline
DSL Spline & 4.3162 & 0.0016 & \cellcolor{white}\textcolor{black}{0.0000} & \cellcolor{gray!50}\textcolor{black}{1.0000} & \cellcolor{gray!50}\textcolor{black}{1.0000} & \cellcolor{white}\textcolor{black}{0.0000} & \cellcolor{white}\textcolor{black}{0.0000} \\
DSL Linear & 4.3162 & 0.0018 & \cellcolor{white}\textcolor{black}{0.0000} & \cellcolor{gray!50}\textcolor{black}{1.0000} & \cellcolor{gray!50}\textcolor{black}{1.0000} & \cellcolor{white}\textcolor{black}{0.0000} & \cellcolor{white}\textcolor{black}{0.0000} \\
DSL Brownian T1 & 4.2599 & 0.0032 & \cellcolor{white}\textcolor{black}{0.0000} & \cellcolor{white}\textcolor{black}{0.0000} & \cellcolor{white}\textcolor{black}{0.0014} & \cellcolor{white}\textcolor{black}{0.0000} & \cellcolor{white}\textcolor{black}{0.0000} \\
DSL Brownian T2 & 5.2945 & 0.0049 & \cellcolor{white}\textcolor{black}{0.0000} & \cellcolor{white}\textcolor{black}{0.0000} & \cellcolor{white}\textcolor{black}{0.0000} & \cellcolor{white}\textcolor{black}{0.0000} & \cellcolor{white}\textcolor{black}{0.0000} \\
DSL Brownian T3 & 6.6704 & 0.0076 & \cellcolor{white}\textcolor{black}{0.0000} & \cellcolor{white}\textcolor{black}{0.0000} & \cellcolor{white}\textcolor{black}{0.0000} & \cellcolor{white}\textcolor{black}{0.0000} & \cellcolor{white}\textcolor{black}{0.0000} \\
DSL Normal & 4.2845 & 0.0044 & \cellcolor{white}\textcolor{black}{0.0000} & \cellcolor{white}\textcolor{black}{0.0000} & \cellcolor{white}\textcolor{black}{0.0000} & \cellcolor{white}\textcolor{black}{0.0000} & \cellcolor{white}\textcolor{black}{0.0000} \\
DSL Uniform 3 & 4.3151 & 0.0067 & \cellcolor{white}\textcolor{black}{0.0000} & \cellcolor{white}\textcolor{black}{0.0000} & \cellcolor{white}\textcolor{black}{0.0000} & \cellcolor{white}\textcolor{black}{0.0000} & \cellcolor{white}\textcolor{black}{0.0000} \\
DSL Uniform 5 & 4.3156 & 0.0094 & \cellcolor{white}\textcolor{black}{0.0000} & \cellcolor{white}\textcolor{black}{0.0000} & \cellcolor{white}\textcolor{black}{0.0000} & \cellcolor{white}\textcolor{black}{0.0000} & \cellcolor{white}\textcolor{black}{0.0000} \\

\bottomrule
\end{tabular}
\caption{The probabilities of each strategy's sample in comparison to the random purchasing distribution. The values of $n$ can vary from 1 to 30, the values of $p$ can be 0, 10\%, 25\%, 33\%, or 50\%. The values of The outcomes of these strategies are statistically unlikely (highlighted exceptions excluded) from random purchasing decisions.}
\label{tab:simulation_results_with_probabilities}
\end{table*}

\begin{table}[t]
\centering
\begin{tabular}{l|cc}
\toprule
Simulation & < 50\% & < 95\% \\
\hline
REG Spline & 100 \% & 80\% \\
DSL Spline & 100 \% & 80\% \\
REG Linear & 97.14\% & 71.43\% \\
DSL Linear & 94.29\% & 74.29\%
\end{tabular}
\caption{The proportions of the 35 price-independent strategies that beat 50\% and 95\% of the random purchase trials.}
\label{tab:fill_at}
\end{table}

The results are best broken up into three groups based on the datasets: curve fitting,  distributional, and Brownian. No strategy was effective for all types because the Brownian datasets had no autocorrelation, meaning that historical prices offered no real information on future prices. For the same reasons, regular purchasing did not prove effective within the random walk price model. Our work suggests that higher autocorrelation is likely important for 

\subsubsection{Spline \& Linear Datasets} These curve fitting datasets comprise a smoothing spline and linear interpolation of REG and DSL. Across these datasets, price-independent or "fill-at" strategies demonstrated the strongest performance, particularly those using levels of 33\% and 10\% with lags of 6 days or 5 days. These levels correspond to 7- or 9-day purchasing intervals, respectively, and the lags favor the end of the regular intervals. This makes sense because the prices in these simulations can experience turning points every 7 days. The extreme values for these functions that are differentiable almost everywhere happen at the ends of the intervals. Given the downward trend in \cite{AAA2025}, favoring the end of the weeks will generally be preferred. Still, these price-independent strategies are preferred regardless of lag as Table \ref{tab:fill_at} demonstrates: 80\% of the price-independent strategies beat 95\% or more of the random purchasing strategies. Some mixed strategies performed well but only at the 10\% tank level meaning that purchases were either made on the 9th or 10th day. So as mixed strategies, they are closer to the price-independent strategies than the price-dependent ones. 

As a whole, mixed strategies did significantly {\it worse} than random purchasing with their mean values being significantly higher than most or all of the random purchasing simulations. The price-dependent strategies generally performed on par with random purchasing, as indicated in Table \ref{tab:simulation_results_with_probabilities}, but with significantly wider distributions. This would suggest that even here, random purchasing would be preferred for household budgets to remain stable.

\subsubsection{Distributional Datasets} All strategies witnessed statistically significant results and all strategies saw a lower mean than the random purchasing distribution. The best performing strategies were the price-dependent strategies:
\begin{itemize}
    \item {\bf Normal Distribution Draws}: Strategies that purchased fuel when the price was lower than the last $n$ days performed best, where for all values of $n$ beat random purchasing. 
    \item {\bf Uniform Distribution Draws}: Strategies that purchased fuel when the price was lower than the $n$-day average performed best, where all values of $n$ beat random purchasing.
\end{itemize}
Mixed strategies also performed well for these datasets, but had much larger ranges of values. Some of the mixed strategies with certain lags performed at levels worse than most random purchasing strategies. For this reason, they were not seen as performing as strongly as the price-dependent strategies. 

\subsubsection{Brownian Datasets} These datasets are arguably the most consequential of the three groups, because they isolate the role of \emph{predictability} from the role of volatility. The Brownian trials occupy the middle of the volatility spectrum in Table~\ref{tab:spectrum}: T1 and T2 bracket the empirical weekly volatilities of REG ($0.0172$) and DSL ($0.0135$), while T3 exceeds them. Despite sitting between the near-deterministic curve-fitting datasets and the highly dispersed distributional datasets, the Brownian group admits \emph{no} consistently effective strategy. As summarized in Table~\ref{tab:performance}, not a single price-independent, price-dependent, or mixed strategy beat $95\%$ of random purchasing across all six Brownian simulations. Table~\ref{tab:simulation_results_with_probabilities} tells the same story: the strategy categories are statistically indistinguishable from---or worse than---the random-purchasing benchmark in essentially every cell. The lone apparent exception, the ``lower than the previous $n$ days'' rule on REG Brownian T1, does not survive replication: the same rule collapses to the benchmark on T2 and T3. Because the three Brownian trials are independent realizations of the same generating process, an edge that appears in one realization but vanishes in the others is the signature of sampling noise around a true null effect rather than a genuine exploitable pattern. Random purchasing is therefore, on average, as good as any history-based rule under the random-walk price model, and we treat the Brownian group as the pivotal counterexample to any volatility-based account of strategy efficacy. We return to why this should be expected in the \nameref{sec:discussion}.
 
 
\section{Discussion}\label{sec:discussion}
 
The three price groups give three qualitatively different answers to the household's question, and the ordering of those answers is not what one would predict from volatility alone. Reading Table~\ref{tab:spectrum} from top to bottom, statistical volatility rises monotonically from the smoothing spline to the five-week uniform sample. The best-performing strategy, however, does not move monotonically with it: price-independent (need-based) rules dominate at the low-volatility end, price-dependent rules dominate at the high-volatility end, and the intermediate-volatility Brownian datasets reward no rule at all. A single explanatory variable---the magnitude of price fluctuations---cannot produce a non-monotonic pattern of this kind. What does organize the results cleanly is the \emph{temporal structure} of each price process: whether, and how, recent prices carry information about future prices.
 
\subsection{Volatility Is Not the Operative Variable}
 
If volatility were the operative variable, the Brownian datasets---whose volatilities ($0.0080$--$0.0210$ for REG and $0.0092$--$0.0154$ for DSL) straddle the empirical weekly values of $0.0172$ and $0.0135$---would be expected to support strategies of intermediate strength, somewhere between the near-deterministic spline and the highly dispersed uniform samples. Instead they support none. The difference between the groups is therefore not one of degree but of kind. The spline, the i.i.d.\ samples, and the random walk are not three points on a single volatility continuum; they are three distinct stochastic processes, and the feature that distinguishes them---how the conditional distribution of tomorrow's price depends on the recent past---is precisely the feature a history-based purchasing rule must exploit. Volatility measures how far prices move; it says nothing about whether those movements are predictable, and predictability is what a strategy converts into savings.
 
\subsection{Three Processes, Three Optimal Responses}
 
\paragraph{Deterministic trend (spline and linear).} A smoothing spline and a linear interpolation are, up to weekly kinks, deterministic functions of time. Serial dependence is near-perfect: given the present price, the next price is fixed to within the curvature of the fit, so the conditional variance of the future price is effectively zero. When the future is (almost) certain and, over the window of \cite{AAA2025}, trending downward, cost minimization collapses to a timing problem with a known answer. For any fixed purchase frequency, the average price paid is minimized by deferring each purchase as long as the model permits---that is, by refilling only when the tank reaches its lowest allowed level. This is exactly the need-based ``fill-at'' rule, which is why $80\%$ of price-independent strategies beat $95\%$ of random purchasing (Table~\ref{tab:fill_at}), while price comparisons to recent history add nothing. In a monotone decline every recent price exceeds the present one, so a ``buy below the $n$-day average'' trigger fires early and systematically pays the trend's higher, earlier prices. The small role of the lag---favoring the sixth day of a seven-day cycle---is the same logic operating within each weekly segment, capturing the local minimum at the end of the interval.
 
\paragraph{Stationary noise (normal and uniform).} In the distributional datasets, consecutive prices are independent draws from a locally stationary distribution: there is no momentum to ride, but there is a stable marginal distribution with a well-defined mean and spread. Here the exploitable quantity is not the trend but the dispersion. Within any window of $n$ recent draws, some fall well below the local mean, and a threshold rule that buys at the $n$-day low or below the $n$-day average is, in effect, harvesting the lower order statistics of that window. If the recent prices are approximately i.i.d.\ with mean $\mu$ and standard deviation $\sigma$, the expected value of the window minimum can be written
\begin{equation}
    \mathbb{E}\!\left[\min_{1\le i \le n} P_i\right] = \mu - \sigma\, a_n,
    \label{eq:order_stat}
\end{equation}
where $a_n$ is a positive constant that increases with the window length $n$. The expected saving relative to a typical purchase is thus proportional to $\sigma$: the wider the price distribution, the more a low-tail draw is worth waiting for. This is why price-dependent strategies are strongest precisely where volatility is highest, and why the advantage is larger for the high-dispersion uniform samples than for the narrower normal ones. It also accounts for the division of labor reported in Section~\ref{sec:analysis}. A plausible reading is that the ``$n$-day low'' rule, which targets the single best recent draw, suits the thin-tailed normal sampling, where the occasional draw far below the mean is the prize; whereas the ``$n$-day average'' rule, which responds to the whole recent window, suits the flat-bodied uniform sampling, where a large and bounded fraction of draws sit meaningfully below the midpoint.
 
\paragraph{Random walk (Brownian).} A Brownian price path has independent increments, which makes it a martingale: absent drift, the best forecast of any future price given the entire history is simply the current price, $\mathbb{E}[P_{t+k}\mid \mathcal{F}_t] = P_t$. The level is highly persistent---today is close to yesterday---so the path can superficially resemble the autocorrelated real series, but the increments carry no information, the process has no mean to revert to, and there is no fixed marginal distribution to define what counts as ``low.'' Both mechanisms that worked elsewhere fail at once: there is no persistent trend for a need-based rule to lean on, and no stationary spread for a threshold rule to harvest. Any timing rule measurable with respect to past prices therefore has zero expected edge over random timing, which is exactly what Table~\ref{tab:performance} reports. Even when a drift term is present, $\mathbb{E}[P_{t+k}\mid \mathcal{F}_t] = P_t + k\mu$, the only response it rewards is to buy uniformly earlier or later---collapsing once more to a need-based timing choice rather than a price-comparison rule. The non-replicating ``win'' on REG Brownian T1 underscores the point: under a true null edge, apparent successes should surface in some realizations and disappear in others, which is precisely what we observe across the three independent trials. This is why the Brownian group, despite mid-range volatility, is the decisive case: it holds dispersion roughly fixed while removing exploitable structure, and the savings vanish.
 
\subsection{Reconciliation with the Empirical Series}
 
These regimes also clarify which model the actual market most resembles. The weekly REG and DSL series are strongly and persistently autocorrelated---Durbin--Watson statistics of $0.025$ and $0.011$, Ljung--Box $p$-values below $10^{-170}$, and significant autoregressive terms out to five and seven weeks (Table~\ref{tab:coefficients})---and they are nonstationary with a pronounced trend over the sample window. The empirical data thus sit far from the random-walk regime and close to the trending, serially dependent regime captured by the spline and linear models. This is reassuring for households: the price model that best matches reality is also the one in which the simplest rule works. A need-based heuristic---refill when the tank is low, with a mild preference for the end of the weekly price cycle---requires no price tracking, is consistent with the bounded-rationality justification of Section~\ref{sec:strategies}, and is exactly the strategy our analysis finds robust under realistic, trending, autocorrelated prices. Chasing recent lows, by contrast, offers households little once prices trend rather than oscillate around a fixed mean.
 
\subsection{Implications for Policy}
 
The same regime distinction reframes the policy question. Because efficacy is governed by the temporal structure of prices rather than by their volatility, interventions that change \emph{how} prices evolve from one period to the next can change which household strategies pay off, independently of the price level. A market pushed toward random-walk behavior---weaker autocorrelation, less mean reversion---erases the value of every simple strategy and leaves households no better than chance, even at unchanged volatility. A market with stable, predictable structure lets a costless need-based rule deliver consistent savings. This supports the paper's central claim. Because gasoline savings propagate into broader consumption and inflation expectations (\cite{Gelman2016}; \cite{Aidala2024}), policymakers concerned with household welfare have reason to attend not only to the retail price level but to the day-to-day dynamics that determine whether households can act on that level at all.
 
\subsection{Limitations}
 
Several caveats bound these conclusions. The reference distribution is a mean of $1000$ random-purchase trials and is approximately normal by the Central Limit Theorem, confirmed by the Shapiro--Wilk and Kolmogorov--Smirnov tests, so the ``beats $95\%$ of random'' criterion functions as a one-sided test at $\alpha = 0.05$ against that benchmark; it identifies strategies that are statistically, not necessarily economically, superior, and the realized dollar gaps implied by Table~\ref{tab:stats} are often small. The model also abstracts from station-to-station price dispersion, brand and quality preferences, and the fixed time and travel costs of more frequent stops, each of which would temper the gains attributed to any active strategy. Finally, the Brownian and distributional datasets are deliberately unrealistic instruments for isolating mechanism: their value is diagnostic rather than predictive, and it is the empirical comparison above---not the synthetic extremes---that anchors the recommendations to observed prices.


\bibliographystyle{plain} 
\bibliography{references}


\onecolumn



\end{document}